\documentclass[a4paper,12pt]{article}
\usepackage{amsmath}
\makeatletter
\def\thebibliography#1{\section{\refname\@mkboth
 {\uppercase{\refname}}{\uppercase{\refname}}}\list
{\@biblabel{\arabic{enumiv}}}{\settowidth\labelwidth{\@biblabel{#1}}
\leftmargin\labelwidth
  \advance\leftmargin\labelsep
   \usecounter{enumiv}    \let\p@enumiv\@empty
  \def\theenumiv{\arabic{enumiv}}}    \def\newblock{\hskip .11em plus.33em
minus.07em}    \sloppy\clubpenalty4000\widowpenalty4000
  \sfcode`\.=1000\relax}
\makeatother

\begin{document}

\title{Examples of Heun and Mathieu Functions as Solutions of Wave Equations in Curved Spaces }
\author{T.Birkandan$^{*}$ and M. Horta\c{c}su$^{*\dag}$  $^{1}$\\
{\small $^{*}${Department of Physics, Istanbul Technical
University,
Istanbul, Turkey}. }\\
{\small $^{\dag}${Feza G\"{u}rsey Institute, Istanbul, Turkey}.}}
\date{\today}
\maketitle

\begin{abstract}
\noindent  We give examples where the Heun function exists as
solutions of wave equations encountered in general relativity.
As a new example we find that while the Dirac equation written
in the background of
Nutku helicoid metric yields Mathieu functions as its solutions in
four space-time dimensions, the trivial generalization to five
dimensions results in the double confluent Heun function.  We
reduce this solution to the Mathieu function with some
transformations.
\end{abstract}


\bigskip\bigskip\bigskip\bigskip\bigskip\bigskip\bigskip\bigskip\bigskip\bigskip\bigskip\bigskip\bigskip\bigskip\bigskip\bigskip\bigskip\bigskip\bigskip\bigskip\bigskip
\footnotetext[1]{E-mail addresses: hortacsu@itu.edu.tr,
birkandant@itu.edu.tr} \pagebreak

\section{Introduction}

Most of the theoretical physics known today is described by a
rather few number of differential equations. If we study only
linear problems, the different special cases of the hypergeometric
or the confluent hypergeometric equation often suffice to analyze
scores of different phenomena. These are two equations of the
Fuchsian type with three regular singular points and one regular,
one irregular singular point respectively. Both of these equations
have simple recursion relations between two consecutive
coefficients when a power expansion solution is attempted. This
fact gives in many instances sufficient information on the general
behavior of the solution. If the problem is nonlinear, one can
usually fit the equation describing the process to one of the
different forms of the Painlev\'{e} equations \cite{Ince}.

It seems to be a miracle that so diverse phenomena, examples in
potential theory or wave equations with physical applications, can
be described with so few equations. Physicists are lucky since
most of the phenomena in physics of the present can be described
in terms of these rather simple functions. Perhaps, this refers to
a symmetry beyond all these things, like the occurrence of
hypergeometric functions may signal to the presence of conformal
symmetry.

In the linear case, sometimes it is necessary to go to equations
with more singular points. The Heun equation
\cite{heun}\cite{schafke}\cite{ronveaux}, its confluent cases, or
its special cases, Mathieu, Lam\'{e}, Coulomb spheroidal equations
etc., all have additional singular points, either four regular, or
two regular, one irregular, or two irregular. The price you pay is
the fact that in general there exists no recursion relation
between two consecutive coefficients when a power series expansion
is used for the solution. A three or four way recursion relation
is often awkward, and it is not easy to deduce valuable analytical
information from such an expansion. New versions of the computer
package Maple, for example Maple 10, gives  graphical
representations of Heun functions.  Although this is a great help,
it is sometimes awkward to use these functions as potentials in
either wave or Schrodinger equations. This may be a reason why
much less is known about these equations compared to
hypergeometric functions and all the other functions derived from
them.

One encounters Mathieu functions when one uses elliptic
coordinates, instead of circular ones, even in two dimensions
\cite {morse1}. Phenomena described by Heun equations are not
uncommon when one studies problems in atomic physics with certain
potentials \cite{bonorino}which combine different inverse powers
starting from the first up to fourth power or combining the
quadratic potential with inverse powers of two, four and six, etc.
.  They also arise when one studies symmetric double Morse
potentials. Slavyanov and Lay \cite{slavyanov} describe different
physical applications of these equations. Atomic physics problems
like separated double wells, Stark effect \cite
{slav115},\cite{slavyanov}, hydrogen-molecule ion  \cite
{slav131}, \cite{slavyanov} use these forms of these equations.
Many different problems in solid state physics like dislocation
movement in crystalline materials, quantum diffusion of kinks
along dislocations are also solved in terms of these functions
\cite{slavyanov}. The famous Hill equation \cite
{hill},\cite{slavyanov}
 for lunar perigee can be cited for an early application in celestial mechanics.

In general relativity, while solving wave equations, we also
encounter different forms of the Heun equations . Teukolsky \cite
{slav127},\cite{slavyanov} studied the perturbations of the Kerr
metric and found out that they were described by two coupled
singly confluent Heun equations.  Quasi-normal modes of rotational
gravitational singularities were also studied by solving this
system of equations \cite{slav87}, \cite{slavyanov}.

In recent applications they become indispensable when one studies
phenomena in higher dimensions, for example the article by G.
Siopsis \cite{siopsis}, or phenomena using different geometries.
An example of the latter case is seen in the example of wave
equations written in the background of these metrics. For
instance, in four dimensions, we may write wave equations in the
background of 4D Euclidean gravity solutions. For the metric
written in the Eguchi-Hanson instanton \cite{eguchi} background,
the hypergeometric function is sufficient to describe the spinor
field solutions \cite{Nuri, villalba}. One, however, has to use
Mathieu functions to describe even the scalar field in the
background of the Nutku helicoid instanton \cite{yavuz1, yavuz2}
when the separation of variables method is used for the solution.
Schmid et al\cite{schmid} have written a short note describing the
occurrence of these equations in general relativity. Their
examples are the Dirac equation in the Kerr-Newman metric and
static perturbations of the non extremal Reisner-Nordstr\"{o}m
solution. They encounter the Generalised Heun Equation
\cite{schafke}\cite{ronveaux}\cite{maier} while looking for the
solutions in these metrics. Here we see that as the metric becomes
more complicated, one has to solve equations with larger number of
singular points, with no simple recursion relations if one
attempts a series type solution. As a particular case of confluent
Heun equation, Fiziev studied the exact solutions of the
Regge-Wheeler equation \cite{fiziev1}\cite{fiziev2}. One also sees
that if one studies similar phenomena in higher dimensions, unless
the metric is a product of simple ones, one has higher chances of
encountering Heun type equations as in the references given
\cite{manvelyan} \cite{oota}.

Here we want to give further examples to this general behaviour.
Our first example will be the case already studied by Sucu and
Unal \cite{Nuri}. In their paper they study the spinor field in
the background of the Nutku helicoid instanton. They obtain an
exact solution, which, however, can be expanded in terms of
Mathieu functions \cite{Chaos}. At this  point note that their
solution in the background of the Eguchi-Hanson instanton is
expanded in terms of hypergeometric functions.  Taking the next 4D
Euclidean gravity solution in row, the Nutku helicoid instanton
with a two centers \cite{gibbons} \cite{valent}, results in a
function with higher singularity structure. The second example is
the similar equation in five dimensions. Here the metric is
extended to five dimensions with a simple addition of the time
coordinate. In this case we obtain the double confluent Heun
function as the solution, which can be reduced to the Mathieu
function with a variable transformation.

At this point the motive for our paper becomes clear.  Both in
four and five dimensions, in the background metric of the Nutku
helicoid instanton, we can write the solutions of the Dirac
equations in terms of Mathieu functions. There is a catch here,
though. Although we can express the solution in a closed form in
the four dimensional case, as done by Sucu and Unal \cite {Nuri},
this is not possible using the solutions in five dimensions, since
the constants used in the two equations are not the same.

In four dimensions we can also calculate the the Greens function
for this differential equation in closed form following  the steps
in reference [18].  In five dimensions we could not succeed in
summing the the infinite series of the product of Mathieu
functions to express both the solution to the differential
equation and the Greens function of the same differential equation
in a closed form.  This is due to the existence of two different
constants in the two Mathieu functions used in the expansion. What
we wanted to show is that going to one higher dimension, we got
solutions which were more complicated.

We describe our examples in the consecutive sections, first the
case in four, then in five dimensions. We then give the solutions
for the scalar operator. We end with some additional remarks. In
our work we use only the massless field, since taking the massive
field is technically like going one higher dimension, which
complicates the problem.  We instead go to one higher dimension in
an explicit fashion in the consecutive section.

\section{Equations in Four Dimensions}

The Nutku helicoid metric is given as

\begin{eqnarray}
{\normalsize ds}^{2} &=&\frac{1}{\sqrt{1+\frac{a^{2}}{r^{2}}}}%
[dr^{2}+(r^{2}+a^{2})d\theta ^{2}+\left( 1+\frac{a^{2}}{r^{2}}\sin
^{2}\theta \right) dy^{2}  \notag \\
&&-\frac{a^{2}}{r^{2}}\sin 2\theta dydz{\normalsize +}\left( 1+\frac{a^{2}}{%
r^{2}}\cos ^{2}\theta \right) {\normalsize dz}^{2}{\normalsize ]}
.
\end{eqnarray}
where $0<r< \infty$, $0 \leq \theta \leq 2\pi$, $y$ and $z$ are
along the Killing directions and will be taken to be periodic
coordinates on a 2-torus \cite {yavuz2}.
This is an example of a
multi-center metric. This metric reduces to the flat metric if we
take $a=0$.
\begin{equation}
ds^{2}=dr^{2}+r^{2}d\theta ^{2}+dy^{2}+dz^{2}.
\end{equation}
If we make the following transformation
\begin{equation}
r=a\sinh x ,
\end{equation}%
the metric is written as
\begin{eqnarray}
ds^{2} &=&\frac{a^{2}}{2}\sinh 2x(dx^{2}+d\theta ^{2})  \notag \\
&&+\frac{2}{\sinh 2x}[(\sinh ^{2}x+\sin ^{2}\theta )dy^{2} \\
&&-\sin 2\theta dydz+(\sinh ^{2}x+\cos ^{2}\theta )dz^{2}]. \notag
\end{eqnarray}%
We use the NP formalism \cite{np1, np2} in four Euclidean
dimensions \cite{goldblatt1} \cite{goldblatt2} \cite{yavuz3}. To
write the Dirac equation in this formalism, we need the choose the
base vectors  and calculate the spin coefficients, the
differential operators and the $\gamma$ matrices in curved space.

\noindent We take  base vectors
\begin{equation}
e_{a}^{\mu }=\{l^{\mu },\bar{l}^{\mu },m^{\mu },\bar{m}^{\mu }\},
\end{equation}%
to give
\begin{equation}
ds^{2}=l\otimes \bar{l}+\bar{l}\otimes l+m\otimes
\bar{m}+\bar{m}\otimes m . \end{equation}%
The tetrad
\begin{equation}
e^{a}=e_{\nu }^{a}dx^{\nu },
\end{equation}%
satisfies
\begin{equation}
\eta _{ab}=e_{a}^{\mu }e_{b}^{\nu }g_{\mu \nu },
\end{equation}%
where  $\eta _{ab}$ is the flat metric.

\noindent We choose
\begin{equation}
l^{\mu }=\frac{1}{a\sqrt{\sinh 2x}}(1,i,0,0),
\end{equation}%
\begin{equation}
m^{\mu }=\frac{1}{\sqrt{\sinh 2x}}(0,0,\cosh (x-i\theta ),i\sinh
(x-i\theta )),
\end{equation}%
giving the two non zero spin coefficients
\begin{equation}
\epsilon =\bar{\epsilon}=\frac{\cosh (2x)}{a\sinh ^{3/2}2x},
\end{equation}%
\begin{equation}
\sigma =\bar{\sigma}=\frac{2}{a\sinh ^{3/2}2x},
\end{equation}%
The rest of the spin coefficients,
\begin{equation}
\kappa =\nu =\gamma =\alpha =\beta =\pi =\tau =\mu =\lambda =\rho
=0.
\end{equation}

 \noindent
 These expressions give the differential operators
\begin{equation}
D=m^{\mu }\partial _{\mu }=\frac{1}{\sqrt{\sinh 2x}}\left[ \cosh
(x-i\theta )\partial _{y}+i\sinh (x-i\theta )\partial _{z}\right],
\end{equation}%
\begin{equation}
\bar{D}=\bar{m}^{\mu }\partial _{\mu }=\frac{1}{\sqrt{\sinh
2x}}\left[ \cosh (x+i\theta )\partial _{y}-i\sinh (x+i\theta
)\partial _{z}\right],
\end{equation}%
\begin{equation}
\delta =l^{\mu }\partial _{\mu }=\frac{1}{a\sqrt{\sinh
2x}}[\partial _{x}+i\partial _{\theta }],
\end{equation}%
\begin{equation}
\bar{\delta}=\bar{l}^{\mu }\partial _{\mu }=\frac{1}{a\sqrt{\sinh 2x}}%
[\partial _{x}-i\partial _{\theta }].
\end{equation}
\noindent
The massive Dirac equation reads%
\begin{equation}
i\gamma ^{\mu }\nabla _{\mu }\Psi =M\Psi,
\end{equation}
where
\begin{equation}
\nabla _{\mu }=\partial _{\mu }-\Gamma _{\mu } ,
\end{equation}%
\noindent
The $\gamma$ matrices can be written in terms of base
vectors as
\begin{equation}
\gamma ^{\mu }=\sqrt{2}\left(
\begin{array}{llll}
0 & 0 & l^{\mu } & m^{\mu } \\
0 & 0 & -\bar{m}^{\mu } & \bar{l}^{\mu } \\
\bar{l}^{\mu } & -m^{\mu } & 0 & 0 \\
\bar{m}^{\mu } & l^{\mu } & 0 & 0%
\end{array}%
\right),
\end{equation}%
These matrices satisfy
\begin{equation}
\{\gamma ^{\mu },\gamma ^{\nu }\}=2g^{\mu \nu }.
\end{equation}

\noindent
The spin connection is written as
\begin{equation}\Gamma _{\mu
}=\frac{1}{4}\gamma _{;\mu }^{\nu }\gamma _{\nu }.
\end{equation}
\noindent
 In expanded form, these equations read

\begin{gather}
\frac{\sqrt{2}}{a\sqrt{\sinh 2x}}\{(\partial _{x}+i\partial
_{\theta })\Psi
_{3}  \notag \\
+a[\cos (\theta +ix)\partial _{y}+\sin (\theta +ix)\partial _{z}]\Psi _{4}-%
\frac{Ma\sqrt{\sinh 2x}}{\sqrt{2}}\Psi _{1}\}=0 ,
\end{gather}
\begin{gather}
\frac{\sqrt{2}}{a\sqrt{\sinh 2x}}\{(\partial _{x}-i\partial
_{\theta })\Psi
_{4}  \notag \\
-a[\cos (\theta -ix)\partial _{y}+\sin (\theta -ix)\partial _{z}]\Psi _{3}-%
\frac{Ma\sqrt{\sinh 2x}}{\sqrt{2}}\Psi _{2}\}=0 ,
\end{gather}%
\begin{gather}
\frac{\sqrt{2}}{a\sqrt{\sinh 2x}}\{\text{(}\partial _{x}-i\partial _{\theta }%
\text{+}\coth 2x\text{)}\Psi _{1}  \notag \\
-a[\cos (\theta +ix)\partial _{y}+\sin (\theta +ix)\partial _{z}]\Psi _{2}-%
\frac{Ma\sqrt{\sinh 2x}}{\sqrt{2}}\Psi _{3}\}=0 ,
\end{gather}%
\begin{gather}
\frac{\sqrt{2}}{a\sqrt{\sinh 2x}}(\partial _{x}+i\partial _{\theta
}+\coth
2x)\Psi _{2}  \notag \\
+a[\cos (\theta -ix)\partial _{y}+\sin (\theta -ix)\partial _{z}]\Psi _{1}-%
\frac{Ma\sqrt{\sinh 2x}}{\sqrt{2}}\Psi _{4}=0 .
\end{gather}
To simplify  calculations,  we will study the massless case. Then
we see that only \{$\Psi _{1},\Psi _{2} $\} and \{$\Psi _{3},\Psi
_{4}$\} are coupled to each other. If we take
\begin{equation}
\Psi_{i} =e^{i(k_{y}y+k_{z}z)}\mathbf{{\Psi_{i} (x,\theta )}}
\end{equation}%
and make the transformations
\begin{equation}
k_{y}=k\cos \phi ,k_{z}=k\sin \phi ,
\end{equation}%
we get
\begin{equation}
\mathbf{{\Psi _{1}}}=\frac{\sinh [x-i(\theta -\phi )]}{\sqrt{\sinh
2x}}\Psi _{1}
\end{equation}%
and
\begin{equation}
\mathbf{{\Psi _{2}}}=\frac{\sinh [x+i(\theta -\phi )]}{\sqrt{\sinh
2x}}\Psi _{2}.
\end{equation}
Now we have to solve

\begin{equation}
\mathit{{L_{1,2}}}\Psi _{1,2}=\frac{-2}{2ak\sqrt{\sinh 2x}}\left\{
\partial _{xx}+\partial _{\theta \theta
}+\frac{a^{2}k^{2}}{2}\left\{ \cos [2(\theta +\phi )]-\cosh
2x\right\} \right\} \Psi _{1,2}=0,
\end{equation}%
whose solutions can be expressed in terms of Mathieu functions.

\begin{align}
\mathbf{{\Psi _{1}}}& =e^{ik(z\sin \phi +y\cos \phi )}\frac{\sinh
[x-i(\theta -\phi )]}{\sqrt{\sinh 2x}}\times  \notag \\
& \{\left[ Se(\zeta_{1},-\frac{a^{2}k^{2}}{4},-ix)+So(\zeta_{1},-\frac{a^{2}k^{2}}{4}%
,-ix)\right] \\
& \times \left[ Se(\zeta_{1},-\frac{a^{2}k^{2}}{4},\theta +\phi )+So(\zeta_{1},-%
\frac{a^{2}k^{2}}{4},\theta +\phi )\right] \},  \notag
\end{align}

\begin{eqnarray}
\mathbf{{\Psi _{2}}} &=&e^{ik(z\sin \phi +y\cos \phi )}\frac{\sinh
[x+i(\theta -\phi )]}{\sqrt{\sinh 2x}}\times  \notag \\
&&\{\left[ Se(\zeta_{2},-\frac{a^{2}k^{2}}{4},-ix)+So(\zeta_{2},-\frac{a^{2}k^{2}}{4}%
,-ix)\right] \\
&&\times \left[ Se(\zeta_{2},-\frac{a^{2}k^{2}}{4},\theta +\phi )+So(\zeta_{2},-%
\frac{a^{2}k^{2}}{4},\theta +\phi )\right] \}.  \notag
\end{eqnarray}

\bigskip
\noindent
 When similar transformations are done for the other
components we get:

\begin{equation}
\mathit{{L_{3}}}\Psi _{3}=\frac{\cosh[x-i(\theta -\phi
)]}{ak}\left\{
\partial _{xx}+\partial _{\theta \theta }+\frac{a^{2}k^{2}}{2}\left\{ \cos
[2(\theta +\phi )]-\cosh 2x\right\} \right\} \Psi _{3}=0,
\end{equation}

\begin{equation}
\mathit{{L_{4}}} \Psi _{4}=\frac{\cosh[x+i(\theta -\phi
)]}{ak}\left\{
\partial _{xx}+\partial _{\theta \theta }+\frac{a^{2}k^{2}}{2}\left\{ \cos
[2(\theta +\phi )]-\cosh 2x\right\} \right\} \Psi _{4}=0.
\end{equation}

\bigskip
\noindent
 The solutions again can be expressed in terms of Mathieu
functions:

\begin{eqnarray}
\mathbf{{\Psi _{3}}} &=&e^{ik(z\sin \phi +y\cos \phi )}\{\left[ Se(\zeta_{3},-%
\frac{a^{2}k^{2}}{4},-ix)+So(\zeta_{3},-\frac{a^{2}k^{2}}{4},-ix)\right]
\notag
\\
&&\times \left[ Se(\zeta_{3},-\frac{a^{2}k^{2}}{4},\theta +\phi )+So(\zeta_{3},-%
\frac{a^{2}k^{2}}{4},\theta +\phi )\right] \},
\end{eqnarray}

\begin{eqnarray}
\mathbf{{\Psi _{4}}} &=&e^{ik(z\sin \phi +y\cos \phi )}\{\left[ Se(\zeta_{4},-%
\frac{a^{2}k^{2}}{4},-ix)+So(\zeta_{4},-\frac{a^{2}k^{2}}{4},-ix)\right]
\notag
\\
&&\times \left[ Se(\zeta_{4},-\frac{a^{2}k^{2}}{4},\theta +\phi )+So(\zeta_{4},-%
\frac{a^{2}k^{2}}{4},\theta +\phi )\right] \}.
\end{eqnarray}
Here note that
\begin{equation*}
{\left[ Se(\zeta_{4},-\frac{a^{2}k^{2}}{4},-ix)+So(\zeta_{4},-\frac{a^{2}k^{2}}{4}%
,-ix)\right] \notag }
\end{equation*}
can be expressed in terms of modified Mathieu functions with real
arguments.  Here $\zeta_{i}$ are separation constants.

\noindent
 At this point also note that we can get solutions for
$\Psi _{3,4}$ in the plane wave form, which are given as
$exp(ka(sin(\theta -\phi +ix)+sin(\theta -\phi -ix)))$, similar to
the ones given by Sucu and Unal \cite{Nuri}. Here, since we want
to point to the occurrence of Mathieu functions in mathematical
physics, we use the product form. This form is also more useful
when  boundary conditions are imposed on the solution.


\section{Equations in Five Dimensions:}

\bigskip The addition of the time component to the previous metric gives:

\begin{equation}
ds^{2}=-dt^{2}+ds_{4}^{2} ,
\end{equation}%
resulting in the massless Dirac equation as:

\begin{equation}
(\gamma ^{\mu }\partial _{\mu }+\gamma ^{t}\partial _{t}-\gamma
^{\mu }\Gamma _{\mu }-\gamma ^{t}\Gamma _{t})\Psi =0 .
\end{equation}%
Here
\begin{equation}
\Gamma _{t}=0
\end{equation}%
and
\begin{equation}
\gamma ^{t}=\left(
\begin{array}{llll}
i & 0 & 0 & 0 \\
0 & i & 0 & 0 \\
0 & 0 & -i & 0 \\
0 & 0 & 0 & -i%
\end{array}%
\right)
\end{equation}

\bigskip \noindent giving the set of equations,

\begin{gather}
\frac{\sqrt{2}}{a\sqrt{\sinh 2x}}\{(\partial _{x}+i\partial
_{\theta })\Psi
_{3}  \notag \\
+a[\cos (\theta +ix)\partial _{y}+\sin (\theta +ix)\partial _{z}]\Psi _{4}+i%
\frac{a\sqrt{\sinh 2x}}{\sqrt{2}}\partial _{t}\Psi _{1}\}=0,
\end{gather}%
\begin{gather}
\frac{\sqrt{2}}{a\sqrt{\sinh 2x}}\{(\partial _{x}-i\partial
_{\theta })\Psi
_{4}  \notag \\
-a[\cos (\theta -ix)\partial _{y}+\sin (\theta -ix)\partial _{z}]\Psi _{3}+i%
\frac{a\sqrt{\sinh 2x}}{\sqrt{2}}\partial _{t}\Psi _{2}\}=0,
\end{gather}%
\begin{gather}
\frac{\sqrt{2}}{a\sqrt{\sinh 2x}}\{(\partial _{x}-i\partial
_{\theta }+\coth
2x)\Psi _{1}  \notag \\
-a[\cos (\theta +ix)\partial _{y}+\sin (\theta +ix)\partial _{z}]\Psi _{2}-i%
\frac{a\sqrt{\sinh 2x}}{\sqrt{2}}\partial _{t}\Psi _{3}\}=0,
\end{gather}%
\begin{gather}
\frac{\sqrt{2}}{a\sqrt{\sinh 2x}}\{(\partial _{x}+i\partial
_{\theta }+\coth
2x)\Psi _{2}  \notag \\
+a[\cos (\theta -ix)\partial _{y}+\sin (\theta -ix)\partial _{z}]\Psi _{1}-i%
\frac{a\sqrt{\sinh 2x}}{\sqrt{2}}\partial _{t}\Psi _{4}\}=0.
\end{gather}

\bigskip
\noindent If we solve for $\Psi _{1}$ and $\Psi _{2}$ and replace
them in the latter equations, we get two equations which has only
$\Psi _{3}$ and $\Psi _{4\text{ }}$ in them. If we take
\begin{equation}
\Psi_{i} =e^{i(k_{t}t+k_{y}y+k_{z}z)}{\mathbf{\Psi_{i}}} (x,\theta
) ,
\end{equation}%
the resulting equations read:

\begin{equation}
\left\{ \partial _{xx}+\partial _{\theta \theta }+\frac{a^{2}k^{2}}{2}%
\left\{ \cos [2(\theta +\phi )]-\cosh 2x\right\}
+2a^{2}k_{t}^{2}\sinh 2x\right\} \bigskip \Psi _{3,4}=0 .
\end{equation}
If we assume that the result is expressed in the product form
$\Psi _{3}=T_{1}(x)T_{2}(\theta ),$ the angular part is again
expressible in terms of Mathieu functions.
\begin{eqnarray}
T_{2}(\theta ) &=&Se\left[ \eta,-\frac{a^{2}k^{2}}{4},\arccos (\sqrt{\frac{%
1+\cos (\theta +\phi )}{2}})\right]  \notag \\
&&+So\left[ \eta,-\frac{a^{2}k^{2}}{4},\arccos (\sqrt{\frac{1+\cos
(\theta +\phi )}{2}})\right] .
\end{eqnarray}
Here $\eta$ is the seperation constant and the periodicity on the
solution makes it equal to the square of an integer.

\noindent
 The equation for $T_{1}$ reads:

\begin{equation}
\left\{ \partial _{xx}-\frac{a^{2}k^{2}}{2}\cosh
2x+2a^{2}k_{t}^{2}\sinh 2x-\eta\right\} T_{1}=0
\end{equation}
The solution of this equation is expressed in terms of double
confluent Heun functions \cite{ronveaux}.

\begin{eqnarray}
T_{1}(x) &=&HeunD\left[ 0,\frac{a^{2}k^{2}}{2}+\eta,4a^{2}k_{t}^{2},\frac{%
a^{2}k^{2}}{2}-\eta,\tanh x\right]  \notag \\
&&+HeunD\left[ 0,\eta+\frac{a^{2}k^{2}}{2},4a^{2}k_{t}^{2},\frac{a^{2}k^{2}}{2}%
-\eta,\tanh x\right] \\
&&\times \int \frac{-dx}{HeunD\left[ 0,\eta+\frac{a^{2}k^{2}}{2}%
,4a^{2}k_{t}^{2},\frac{a^{2}k^{2}}{2}-\eta,\tanh x\right] ^{2}} .
\notag
\end{eqnarray}

 We only take the first function and discard the second solution.
We see that as $x$ goes to infinity, the function given above
diverges. The function is finite at $x=0$ though.  In order to get
well defined functions, we study the region where $x\leq F$, where
F is a finite value. We will give a way to determine F below.

\noindent
 We can use either Dirichlet or Neumann boundary
conditions for our problem in four dimensions.  There is an
obstruction in odd Euclidean dimensions that makes us use the
Atiyah-Patodi-Singer \cite {Atiyah} spectral boundary conditions.
These boundary conditions can also be used in even Euclidean
dimensions if we want to respect the charge conjugation and the
$\gamma^5$ symmetry \cite{hortac}.

\noindent
 Just to show the differences with the four dimensional
solution, we attempt to write this expression in terms of Mathieu
functions. This can be done after few transformations. We define

\begin{equation}
A=2a^{2}k_{t}^{2},
\end{equation}

\begin{equation}
B=-\eta,
\end{equation}

\begin{equation}
C=-\frac{a^{2}k^{2}}{2},
\end{equation}
and use the transformation

\begin{equation}
z=e^{-2x}.
\end{equation}
Then the differential operator is expressed as

\begin{equation}
\mathit{O}=4z^{2}\partial _{zz}+4z\partial _{z}+A^{\prime }z+B+C^{\prime }%
\frac{1}{z}.
\end{equation}
Here

\begin{equation}
A^{\prime }=\frac{C-A}{2}
\end{equation}

\begin{equation}
C^{\prime }=\frac{C+A}{2}
\end{equation}
If we take
\begin{equation}
\sqrt{\frac{C^{\prime }}{A^{\prime }}}u=z,
\end{equation}
and

\begin{equation}
w=\frac{1}{2}(u+\frac{1}{u})
\end{equation}
and set $E=\sqrt{A^{\prime }C^{\prime }}$ we get,

\begin{equation}
\mathit{O}=(w^{2}-1)\partial _{ww}+w\partial
_{w}+\frac{E}{2}w+\frac{B}{4}.
\end{equation}
The solution of this equation is also expressible in terms of
Mathieu functions given as:


\begin{eqnarray}
R(z) &=& Se(-B,E,\arccos \sqrt{w+1})+So(-B,E,\arccos \sqrt{w+1})
\end{eqnarray}
At this point we  see a natural limitation in the values that can
be taken by our radial variable since the argument of the function
$\arccos$ can not exceed unity.  The fact that $\sqrt{w+1} $ can
not exceed unity limits the values our initial variable $x$ can
take, thus determining F which imposed on our solution in equation
[50].

\noindent
 Here we also see a difference from the four dimensional case.
Although both the radial and the angular part can be written in
terms of Mathieu functions, the constants are different,modified
by the presence of the new $-2a^2 k_{t^2}^2$ term, which makes the
summation of these functions to form the propagator quite
difficult.

\noindent
 In four dimensions we can use the summation
formula \cite{morse, yavuz2} for the product of four Mathieu
functions, two of them for the angular and the other two for the
radial part, summing them to give us a Bessel type expression.
This result makes the calculation of the propagator, similar to
the case given in reference [18] possible.  In that case, the
similar analysis also gives the solution to the differential
equation in a closed form as given in reference [15].
 Here since the radial and the angular parts have
different constants, this summation formula is not applicable
write the Greens function in a closed form. We
 also see that  using the generating function formula  for these
functions to write the solution to the differential equation in
terms of plane waves, as described in the paper by L.Chaos-Cador
et al., is not applicable in the five dimensional case due to the
same reason.

\bigskip

\section{Laplacian}

In this section we give the Laplacian operator written in this
background. It is used for the calculation of the field equation
for the scalar particle,
  similar to the case
studied in reference [18].

\begin{equation}
H:=\frac{1}{\sqrt{-g}}\partial _{\nu }\sqrt{-g}g^{\mu \nu
}\partial _{\mu }
\end{equation}%
\begin{eqnarray}
H &:&=-\partial _{tt}+\partial _{xx}+\partial _{\theta \theta
}+a^{2}\sinh
^{2}2x(\partial _{yy}+\partial _{zz})  \notag \\
&&+a^{2}(\cos \theta \partial _{y}+\sin \theta \partial
_{z})^{2}-a^{2}\sinh x\cosh x\partial _{tt}
\end{eqnarray}%
We see that there are three Killing vectors and one quadratic
Killing tensor with eigenvalues given below. From the Killing
tensor we can construct a second order operator tensor \cite
{yavuz2}
\begin{equation}
K=-\partial _{\theta \theta }-a^{2}(\cos \theta \partial _{y}+\sin
\theta
\partial _{z})^{2}
\end{equation}%
\begin{equation}
K\Phi =\lambda \Phi .
\end{equation}%
We use $\lambda $ as the separation constant. The other
eigenvalues are
\begin{equation}
\partial _{t}\Phi =k_{t}\Phi ,
\end{equation}%
\begin{equation}
\partial _{y}\Phi =k_{y}\Phi ,
\end{equation}%
\begin{equation}
\partial _{z}\Phi =k_{z}\Phi .
\end{equation}%
We have
\begin{equation}
\Phi =e^{i(k_{t}t+k_{y}y+k_{z}z)}R(x)S(\theta )
\end{equation}%
where $S$ obeys the equation
\begin{equation}
\frac{d^{2}\tilde{S}}{d\tilde{\theta}^{2}}+(\lambda -a^{2}k^{2}\cos ^{2}%
\tilde{\theta})\tilde{S}=0.
\end{equation}%
Here
\begin{equation}
\tilde{\theta}=\theta -\phi
\end{equation}%
and the solution reads
\begin{eqnarray}
S(\theta ) &=&Se(\frac{-a^{2}k^{2}}{2}+\lambda ,\frac{a^{2}k^{2}}{%
4},\theta -\phi )  \notag \\
&&+So(\frac{-a^{2}k^{2}}{2}+\lambda ,\frac{a^{2}k^{2}}{4},\theta
-\phi ).
\end{eqnarray}%
$\lambda$ is the separation constant which goes to the square of
an integer due to periodicity of the function.

\noindent
 The radial part
obeys
\begin{equation}
\frac{d^{2}R}{dx^{2}}+a^{2}(\sinh x\cosh xk_{t}^{2}-k^{2}\sinh ^{2}x-\frac{%
\lambda }{a^{2}})R=0.
\end{equation}
The solution to this equation, the radial solution, can be written
in terms of Double confluent Heun functions,

\begin{eqnarray}
R(x) &=&HeunD\left[ 0,a^{2}k^{2}-\lambda,-a^{2}k_{t}^{2},\lambda,\tanh x\right]  \notag \\
&&+HeunD\left[ 0,a^{2}k^{2}-\lambda,-a^{2}k_{t}^{2},\lambda,\tanh x\right] \\
&&\times \int \frac{-dx}{HeunD\left[
0,a^{2}k^{2}-\lambda,-a^{2}k_{t}^{2},\lambda,\tanh x\right] ^{2}}
. \notag
\end{eqnarray}
and this can be reduced to the modified Mathieu function after
performing similar transformation as in the spinor case treated
above. Taking $A=\frac{a^{2}k_{t}^{2}}{2}$,
$B=\frac{a^{2}k^{2}}{2}-\lambda$ and $C=\frac{-a^{2}k^{2}}{2}$,
and use $z=e^{-2x}$ transformation we get the same result as in
equation (61).

\section{Conclusion}

Here we related solutions of the Dirac equation in the background
of the Nutku helicoid solution in five dimensions to the Double
confluent Heun function. This solution can be also expressed in
terms of the Mathieu function, which is more familiar to the
physics community, at the expense of using the $z=e^{-2x}$
transformation, which maps infinity to zero followed by a
rescaling and a further transformation where, aside from the
scaling, we are taking the hyperbolic cosine of the original
radial variable. This transformation also brings a natural limit
on the radial variable.
\noindent
Essentially we do not gain much,
since the Mathieu function also has a three way recursion
relation, and a not very handy generating function, compared to
the generating functions belonging to the hypergeometric family.
Such a transformation to Mathieu functions may not be possible
when more complicated backgrounds are taken. Although in many
cases the Heun function may reduce to more simple forms
\cite{maier2} \cite{gurappa}, there are metrics in general
relativity where this may not be possible. The Nutku helicoid
metric is such an example.

\noindent
 We know that using Maple 10 programme we can see the
graphical representation of these functions.  Different asymptotic
solutions are also studied in references [3,4,5] and in individual
papers. Still it is our feeling that most of the mathematical
physics community is not at ease with these functions as they are
with the better known hypergeometric or confluent hypergeometric
functions. The literature on these functions is also very limited.
We think we have covered most of the monographs where these
functions are thoroughly studied in our bibliography. If we search
the SPIRES web site, which is commonly used by mathematical
physicists, the number of entities is rather small. This is the
reason we think one should be exposed to its applications more
often.

\noindent
 We also wanted to show in our work that when one uses
more complicated forms of similar structures, one gets more
complicated solutions.  What we mean by this phrase that if one
writes the Dirac equation in the background of the simplest
Euclidean solution in four dimensions, the Eguchi-Hanson solution,
one gets the hypergeometric function as solutions of the wave
equation \cite {Nuri}. If we use the next solution in the order of
complexity, the Nutku helicoid solution, one gets a solution of
the Heun type. We think the solution obtained in reference [15]
for this case in terms of exponential functions is somewhat
misleading, since there they get the generating function of the
Mathieu functions.  This may not be recognized by the readers
unless one is an expert in this field.

\noindent
 We also studied the same equation in one higher
dimension. Often increasing the number of dimensions of the
manifold in which the wave equation is written, results in higher
functions as solutions. Here we call a function of a higher type
if it has more singularities. In this respect Heun function
belongs to a higher form than the hypergeometric function.  In the
\emph{Introduction} we gave examples of the use of Heun functions
encountered in different physics problems.

\noindent
 In the future we will try to find further examples of
such functions encountered in solutions to wave equations in
general relativity.

\noindent \linebreak
 \textbf{Acknowledgement}: We thank Profs.
Yavuz Nutku, Alikram N. Aliev and Paul Abbott and Ferhat Ta\c sk\i
n for correspondence, for discussions and both scientific and
technical assistance throughout this work. The work of M.H. is
also supported by TUBA, the Academy of Sciences of Turkey. This
work is also supported by TUBITAK, the Scientific and
Technological Council of Turkey.

\begin{center}
\bigskip
\end{center}

\end{document}